\documentclass[twocolumn,preprintnumbers,prl]{revtex4-1}
\pdfoutput=1
\usepackage{amsmath,amssymb,graphicx,bm,color}

\begin{document}

\title{LHC as an Axion Factory:\\[1mm] 
Probing an Axion Explanation for $\mathbf{(g-2)_\mu}$ with Exotic Higgs Decays}

\preprint{MITP/17-023}
\preprint{April 26, 2017}

\author{Martin Bauer$^a$}
\author{Matthias Neubert$^{b,c}$}
\author{Andrea Thamm$^b$}

\affiliation{$^a$Institut f\"ur Theoretische Physik, Universit\"at Heidelberg, Philosophenweg 16, 69120 Heidelberg, Germany\\
${}^b$PRISMA Cluster of Excellence {\em\&} MITP, Johannes Gutenberg University, 55099 Mainz, Germany\\
${}^c$Department of Physics {\em\&} LEPP, Cornell University, Ithaca, NY 14853, U.S.A.}

\begin{abstract}
We argue that a large region of so far unconstrained parameter space for axion-like particles (ALPs), where their couplings to the Standard Model are of order $(0.01\!-\!1)\,\mbox{TeV}^{-1}$, can be explored by searches for the exotic Higgs decays $h\to Za$ and $h\to aa$ in Run-2 of the LHC. Almost the complete region in which ALPs can explain the anomalous magnetic moment of the muon can be probed by searches for these decays with subsequent decay $a\to\gamma\gamma$, even if the relevant couplings are loop suppressed and the $a\to\gamma\gamma$ branching ratio is less than~1. 
\end{abstract}
\maketitle

Axion-like particles (ALPs) appear in well motivated extensions of the Standard Model (SM), e.g.\ as a way to address the strong CP problem, as mediators between the SM and a hidden sector, or as pseudo Nambu-Goldstone bosons in extensions of the SM with a broken global symmetry. If ALP couplings to muons and photons are present, the $3.6\sigma$ deviation of the anomalous magnetic moment of the muon $a_\mu=(g-2)_\mu/2$ from its SM value can be explained by ALP exchange \cite{Chang:2000ii,Marciano:2016yhf}. Collider experiments can be used to search directly and indirectly for ALPs. Besides ALP production in association with photons, jets and electroweak gauge bosons \cite{Kleban:2005rj,Mimasu:2014nea,Jaeckel:2015jla,Brivio:2017ije}, searches for the decay $Z\to\gamma a$ are sensitive to ALPs with up to weak-scale masses \cite{Kim:1989xj,Djouadi:1990ms,Rupak:1995kg}. Utilizing the exotic Higgs decay $h\to aa$ to search for light pseudoscalars was proposed in \cite{Dobrescu:2000jt,Dobrescu:2000yn,Chang:2006bw}. Several experimental searches for this mode have been performed, constraining various final states \cite{Chatrchyan:2012cg,CMS:2015iga,CMS:2016cel,Aad:2015bua,Khachatryan:2015nba,CMS:2016tgd,Khachatryan:2017mnf}. Surprisingly, the related decay $h\to Za$ has not been studied experimentally, even though analogous searches for new heavy scalar bosons decaying into $Za$ have been performed \cite{Khachatryan:2016are}. The reason is, perhaps, the suppression of the $h\to Za$ decay in the decoupling limit in two-Higgs-doublet models in general and supersymmetric models in particular \cite{Branco:2011iw}. In models featuring a gauge-singlet ALP, there is no dimension-5 operator mediating $h\to Za$ decay at tree level, and hence this mode has not received much theoretical attention either. Here we point out that fermion-loop graphs arising at dimension-5 order and tree-level contributions of dimension-7 operators can naturally induce a $h\to Za$ decay rate of similar magnitude as the $h\to Z\gamma$ decay rate in the SM, which is a prime target for Run-2 at the LHC. Furthermore, in certain classes of UV completions the $h\to Z\gamma$ branching ratio can be enhanced parametrically to the level of ${\cal O}(10\%)$ and higher. A search for this decay mode is therefore well motivated and can provide non-trivial information about the underlying UV theory. 

In this letter we show that searches for $h\to Za$ and $h\to aa$ decays in Run-2 at the LHC can probe a large region of so far unconstrained parameter space in the planes spanned by the ALP mass and its couplings to photons or leptons, covering in particular the difficult region above 30\,MeV and probing ALP--photon couplings as small as $10^{-10}\,\mbox{GeV}^{-1}$. If the $(g-2)_\mu$ anomaly is explained by a light pseudoscalar, this particle will be copiously produced in Higgs decays and should be discovered at the LHC. A detailed discussion of the searches presented here, along with a comprehensive analysis of electroweak precision bounds, flavor constraints and the relevance of ALPs to other low-energy anomalies, will be presented elsewhere \cite{inprep}.

We consider a light, gauge-singlet CP-odd boson $a$, whose mass is protected by a (approximate) shift symmetry. Its interactions with SM fermions and gauge fields start at dimension-5 order and are described by the effective Lagrangian \cite{Georgi:1986df} 
\begin{align}
   {\cal L}_{\rm eff}
   &= g_s^2\,C_{GG}\,\frac{a}{\Lambda}\,G_{\mu\nu}^A\,\tilde G^{\mu\nu,A} 
    + g^2\,C_{WW}\,\frac{a}{\Lambda}\,W_{\mu\nu}^A\,\tilde W^{\mu\nu,A} \notag\\
   &\mbox{}+ g^{\prime\,2}\,C_{BB}\,\frac{a}{\Lambda}\,B_{\mu\nu}\,\tilde B^{\mu\nu}
    + \sum_f \frac{c_{ff}}{2}\,\frac{\partial^\mu a}{\Lambda}\,\bar f\gamma_\mu\gamma_5 f \,. 
\end{align}
Here $\Lambda$ is the characteristic scale of global symmetry breaking (often called $f_a$ in the literature on ALPs), which we assume to be above the weak scale. We neglect flavor off-diagonal couplings, which will play no role for our analysis. After electroweak symmetry breaking, the effective ALP coupling to two photons is described by a term analogous to the hypercharge coupling, but with gauge coupling $e^2$ and coefficient $C_{\gamma\gamma}=C_{WW}+C_{BB}$. Note that at this order there are no ALP couplings to the Higgs doublet $\phi$. They appear first at dimension-6 and 7 and are given by 
\begin{equation}\label{LeffD>5}
\begin{aligned}
   {\cal L}_{\rm eff}^{D\ge 6}
   &= \frac{C_{ah}}{\Lambda^2} \left( \partial_\mu a\right)\!\left( \partial^\mu a\right) \phi^\dagger\phi \\
   &\quad\mbox{}+ \frac{C_{Zh}^{(7)}}{\Lambda^3} \left( \partial^\mu a\right) 
    \left( \phi^\dagger\,iD_\mu\,\phi + \mbox{h.c.} \right) \phi^\dagger\phi + \dots \,.
\end{aligned}
\end{equation}
The first term is the leading Higgs-portal interaction allowed by the shift symmetry, while the second term is the leading polynomial operator mediating the decay $h\to Za$ at tree level \cite{Bauer:2016ydr}. If the electroweak symmetry is realized non-linearly, insertions of $\phi^\dagger\phi$ are accompanied by factors $1/f^2$ rather than $1/\Lambda^2$, where $f$ is the analog of the pion decay constant \cite{Feruglio:1992wf}. As a result, the contribution of $C_{Zh}^{(7)}$ can be enhanced by a factor $\sim\Lambda^2/f^2$ if $f<\Lambda$ \cite{Brivio:2017ije}. Importantly, in models featuring heavy particles which receive their mass from electroweak symmetry breaking, an additional non-polynomial dimension-5 operator
\begin{equation}\label{Leffnonpol} 
   {\cal L}_{\rm eff}^{\rm non-pol} 
   = \frac{C_{Zh}^{(5)}}{\Lambda} \left( \partial^\mu a\right)
    \left( \phi^\dagger\,iD_\mu\,\phi + \mbox{h.c.} \right) \ln\frac{\phi^\dagger\phi}{\mu^2} + \dots
\end{equation}
can be generated \cite{Bauer:2016ydr}. It gives a contribution to the $h\to Za$ amplitude that is parametrically enhanced compared with the $h\to aa$ amplitude. The decay $h\to Za$ is unique in the sense that a tree-level dimension-5 coupling can {\em only\/} arise from a non-polynomial operator. 
A search for this decay mode can thus provide complementary information to $h\to aa$ searches and offer important clues about the underlying UV theory. 

\begin{figure}
\begin{center}
\includegraphics[width=0.38\textwidth]{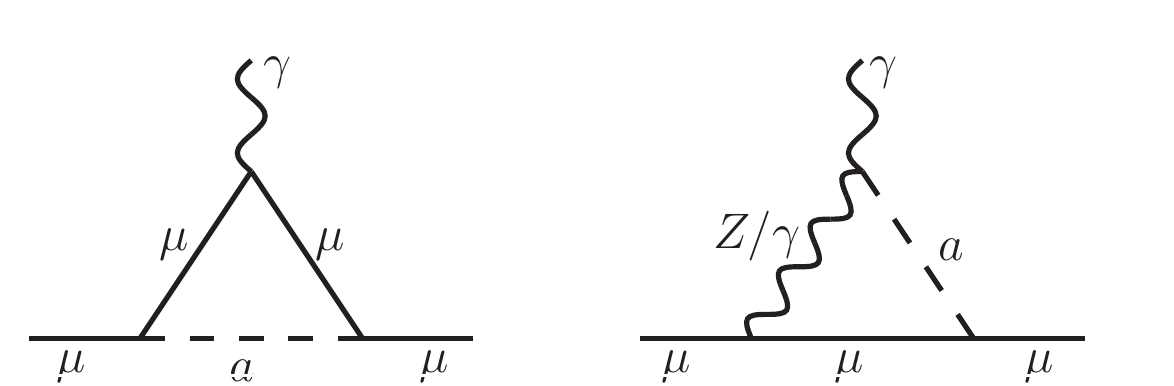}
\end{center}
\vspace{-3mm}
\caption{\label{fig:amugraphs} ALP-induced one-loop contributions to the anomalous magnetic moment of the muon.}
\end{figure}

\begin{figure}
\begin{center}
\includegraphics[width=0.45\textwidth]{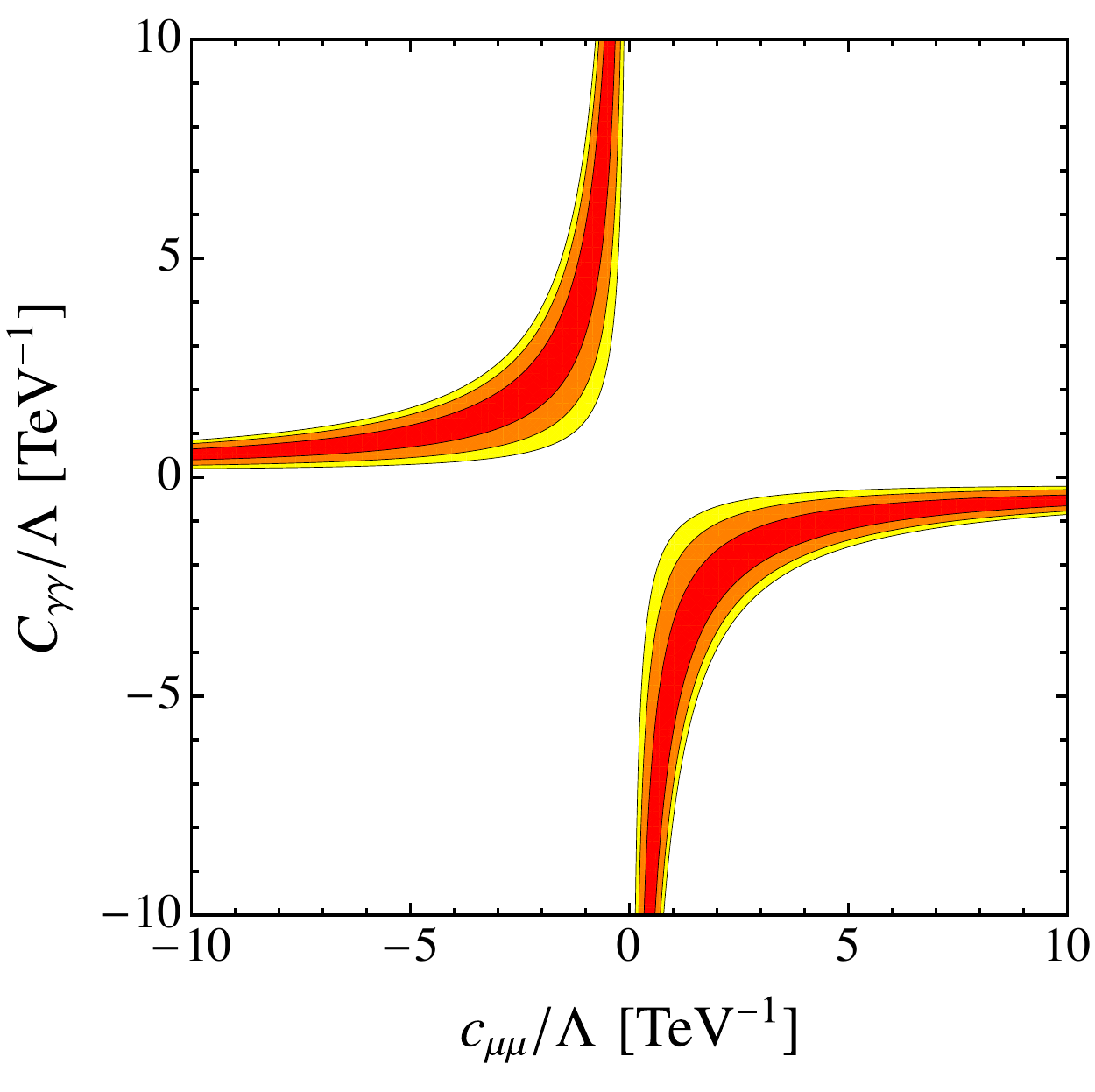}
\end{center}
\vspace{-4mm}
\caption{\label{fig:amu} Regions in ALP coupling space where the experimental value of $(g-2)_\mu$ is reproduced at 68\% (red), 95\% (orange) and 99\% (yellow) confidence level (CL), for $m_a=1$\,GeV.}
\end{figure}

In this letter we consider decays of the ALP into photons and charged leptons, with decay rates given by
\begin{equation}
\begin{aligned}
   \Gamma(a\to\gamma\gamma) 
   &= \frac{4\pi\alpha^2 m_a^3}{\Lambda^2}\,C_{\gamma\gamma}^2 \,, \\
   \Gamma(a\to\ell^+\ell^-) 
   &= \frac{m_a m_\ell^2}{8\pi\Lambda^2}\,c_{\ell\ell}^2\,\sqrt{1-\frac{4m_\ell^2}{m_a^2}} \,.
\end{aligned}
\end{equation}
The same couplings enter the diagrams shown in Figure~\ref{fig:amugraphs}, which show the ALP-induced contributions to the anomalous magnetic moment $a_\mu$ of the muon, whose experimental value differs by more than 3$\sigma$ from the SM prediction: $a_\mu^{\rm exp}-a_\mu^{\rm SM}=(288\pm 63\pm 49)\cdot 10^{-11}$ \cite{Olive:2016xmw}. It has been emphasized recently that this discrepancy can be explained by postulating the existence of an ALP with sizeable couplings to both photons and muons \cite{Marciano:2016yhf}. While the first graph in Figure~\ref{fig:amugraphs} gives a contribution of the wrong sign \cite{Leveille:1977rc,Haber:1978jt}, the second diagram can overcome this contribution if the Wilson coefficient $C_{\gamma\gamma}$ is sufficiently large \cite{Chang:2000ii,Marciano:2016yhf}. At one-loop order, we find the new-physics contribution 
\begin{equation}\label{deltaamu}
   \delta a_\mu \!=\! \frac{m_\mu^2}{\Lambda^2}\,\bigg\{ - \frac{c_{\mu\mu}^2}{16\pi^2}\,h_1(x) 
    - \frac{2\alpha}{\pi}\,c_{\mu\mu} C_{\gamma\gamma} \bigg[ \ln\frac{\Lambda^2}{m_\mu^2} 
    - h_2(x) \bigg] \bigg\} ,
\end{equation}
where $x=m_a^2/m_\mu^2$. The functions $h_i(x)$ are positive and satisfy $h_1(0)=h_2(0)=1$ and $h_1(x)\approx 0$, $h_2(x)\approx(\ln x+\frac32)$ for $x\gg 1$. Their analytical expressions will be given in \cite{inprep}. Our result for the logarithmically enhanced contribution proportional to $C_{\gamma\gamma}$ agrees with \cite{Marciano:2016yhf}. We omit the numerically subdominant contribution from $Z$ exchange, which is suppressed by $(1-4\sin^2\theta_w)$ and comes with a smaller logarithm $\ln(\Lambda^2/m_Z^2)$. A positive shift of $a_\mu$ can be obtained if $c_{\mu\mu}$ and $C_{\gamma\gamma}$ have opposite signs. Figure~\ref{fig:amu} shows the parameter space in the $c_{\mu\mu}-C_{\gamma\gamma}$ plane in which the muon anomaly can be explained in terms of an ALP with mass of 1\,GeV (we use $\Lambda=1$\,TeV in the argument of the logarithm). The contours are insensitive to $m_a$ for lighter ALP masses and broaden slightly for $m_a>1$\,GeV. A resolution of the anomaly is possible without much tuning as long as one of the two coefficients is of order $\Lambda/\mbox{TeV}$, while the other can be of similar order or larger. Since $c_{\mu\mu}$ enters observables always in combination with $m_\mu$, it is less constrained by perturbativity than $C_{\gamma\gamma}$. We thus consider the region where $|C_{\gamma\gamma}|/\Lambda\lesssim 2\,\mbox{TeV}^{-1}$ and $|c_{\mu\mu}|\ge|C_{\gamma\gamma}|$ as the most plausible parameter space.

We now turn our attention to the exotic Higgs decays $h\to Za$ and $h\to aa$, arguing that over wide regions of parameter space -- including the region motivated by $(g-2)_\mu$ -- the high-luminosity LHC can serve as an ALP factory. At tree-level, the effective interactions in (\ref{LeffD>5}) and (\ref{Leffnonpol}) yield the decay rates
\begin{equation}\label{GammaSZh7}
\begin{aligned}
   \Gamma(h\to Za) 
   &= \frac{m_h^3}{16\pi\Lambda^2}\,C_{Zh}^2\,\lambda^{3/2}
    \bigg(\frac{m_Z^2}{m_h^2},\frac{m_a^2}{m_h^2}\bigg) \,, \\
   \Gamma(h\to aa) 
   &= \frac{v^2 m_h^3}{32\pi\Lambda^4}\,C_{ah}^2 \bigg( 1-\frac{2m_a^2}{m_h^2} \bigg)^2
    \sqrt{1-\frac{4m_a^2}{m_h^2}} \,,
\end{aligned}
\end{equation}
where $\lambda(x,y)=(1-x-y)^2-4xy$, and we have defined $C_{Zh}\equiv C_{Zh}^{(5)}+\frac{v^2}{2\Lambda^2}\,C_{Zh}^{(7)}$. Integrating out the top-quark yields the one-loop contributions $\delta C_{Zh}\approx -0.016\,c_{tt}$ and $\delta C_{ah}\approx 0.173\,c_{tt}^2$ \cite{inprep}. For natural values of the Wilson coefficients the rates in (\ref{GammaSZh7}) can give rise to large branching ratios. For instance, one finds $\mbox{Br}(h\to Za)=0.1$ for $|C_{Zh}|/\Lambda\approx 0.34\,\mbox{TeV}^{-1}$ and $\mbox{Br}(h\to aa)=0.1$ for $|C_{ah}|/\Lambda^2\approx 0.62\,\mbox{TeV}^{-2}$. Even in the absence of large tree-level contributions, the loop-induced top-quark contribution yields $\mbox{Br}(h\to aa)=0.01$ for $|c_{tt}|/\Lambda\approx 1.04\,\mbox{TeV}^{-1}$, while a combination of the top-quark contribution and the dimension-7 contribution from $C_{Zh}^{(7)}$ can give $\mbox{Br}(h\to Za)={\cal O}(10^{-3})$ without tuning. With such rates, large samples of ALPs will be produced in Run-2 of the LHC. The model-independent bound $\text{Br}(h\to \text{BSM})<0.34$ derived from the global analysis of Higgs couplings \cite{Khachatryan:2016vau} implies $|C_{Zh}|/\Lambda\lesssim 0.72\,\mbox{TeV}^{-1}$ and $|C_{ah}|/\Lambda^2<1.34\,\mbox{TeV}^{-2}$ at 95\% CL. 

\begin{figure}
\begin{center}
\includegraphics[width=0.5\textwidth]{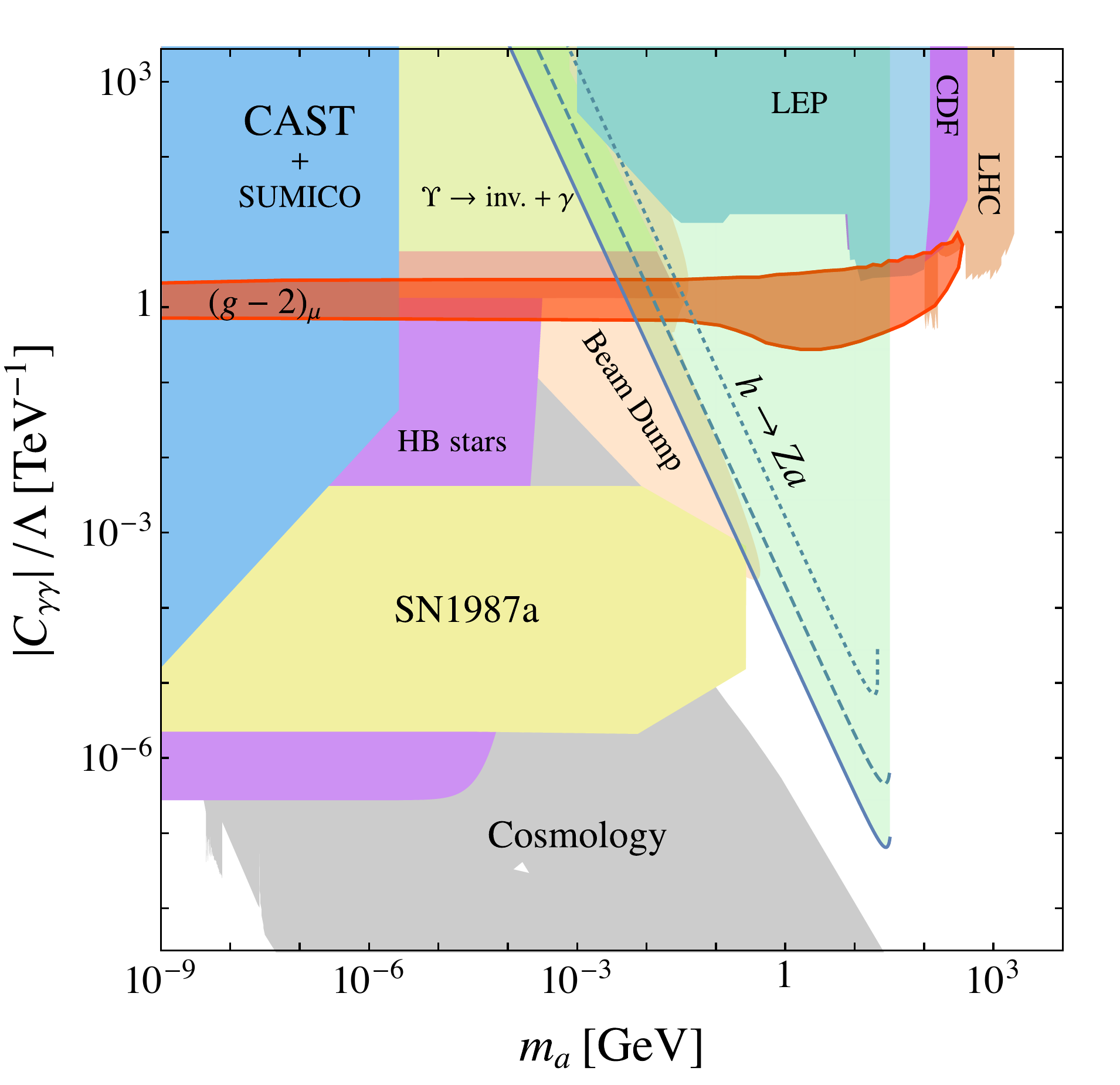}\quad\includegraphics[width=0.5\textwidth]{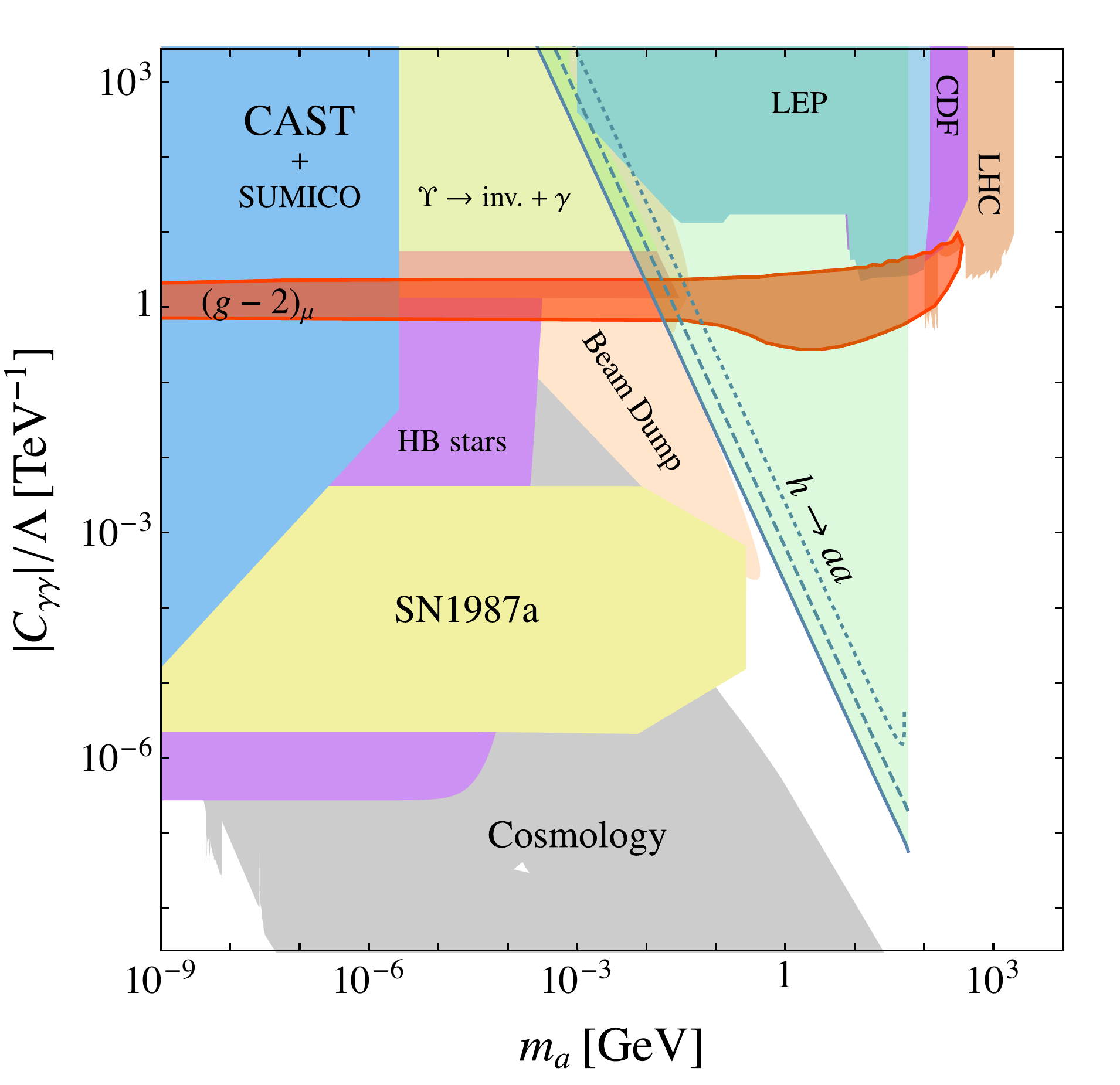}
\end{center}
\vspace{-3mm}
\caption{\label{fig:CggexclusionAA} Constraints on the ALP mass and coupling to photons derived from various experiments (colored areas without boundaries, adapted from \cite{Jaeckel:2015jla}) along with the parameter regions (shaded in light green) that can be probed in LHC Run-2 (300\,fb$^{-1}$ integrated luminosity) using the Higgs decays $h\to Za\to\ell^+\ell^-\gamma\gamma$ (top) and $h\to aa\to 4\gamma$ (bottom). We require at least 100 signal events in each channel. The contours in the upper panel correspond to $|C_{Zh}|/\Lambda=0.72\,\mbox{TeV}^{-1}$ (solid), $0.1\,\mbox{TeV}^{-1}$ (dashed) and $0.015\,\mbox{TeV}^{-1}$ (dotted). Those in the lower panel refer to $|C_{ah}|/\Lambda^2=1\,\mbox{TeV}^{-2}$ (solid), $0.1\,\mbox{TeV}^{-2}$ (dashed) and $0.01\,\mbox{TeV}^{-2}$ (dotted). The red band shows the preferred parameter space where the $(g-2)_\mu$ anomaly can be explained at 95\% CL.}
\end{figure}

If the ALP is light or weakly coupled to SM fields, its decay length can become macroscopic, and hence only a small fraction of ALPs decay inside the detector. Since to good approximation Higgs bosons at the LHC are produced along the beam direction, the {\em average\/} decay length of the ALP perpendicular to the beam is $L_a^\perp(\theta)=\sin\theta\,\beta_a\gamma_a/\Gamma_a$, where $\theta$ is the angle of the ALP with respect to the beam axis in the Higgs-boson rest frame, $\beta_a$ and $\gamma_a$ are the usual relativistic factors in that frame, and $\Gamma_a$ is the total decay width of the ALP. If the ALP is observed in the decay mode $a\to X\bar X$, we can express its total width in terms of the branching fraction and partial width for this decay, i.e.\
\begin{equation}
   L_a^\perp(\theta) = \sin\theta \sqrt{\gamma_a^2-1}\,\,\frac{\mbox{Br}(a\to X\bar X)}{\Gamma(a\to X\bar X)} \,.
\end{equation}
The boost factor is $\gamma_a=(m_h^2-m_Z^2+m_a^2)/(2m_a m_h)$ for $h\to Za$ and $\gamma_a=m_h/(2m_a)$ for $h\to aa$. As a consequence, only a fraction of events given by 
\begin{equation}\label{fdecay}
   f_{\rm dec} = 1 - \left\langle e^{-L_{\rm det}/L_a^\perp(\theta)} \right\rangle \,,
\end{equation}
where the brackets mean an average over solid angle, decays before the ALP has traveled a distance $L_{\rm det}$ set by the relevant detector components. We define the effective branching ratios
\begin{align}\label{eq:effBR}
   &\mbox{Br}(h\to Za\to\ell^+\ell^- X\bar X) \big|_{\rm eff} = \mbox{Br}(h\to Za) \notag\\
   &\hspace{15mm}\times \mbox{Br}(a\to X\bar X)\,f_{\rm dec}\,\mbox{Br}(Z\to \ell^+\ell^-) \,, \\
   &\mbox{Br}(h\to aa\to 4X) \big|_{\rm eff} 
    = \mbox{Br}(h\to aa)\,\mbox{Br}(a\to X\bar X)^2\,f_{\rm dec}^2 \,, \notag
\end{align}
where $\mbox{Br}(Z\to \ell^+\ell^-)=0.0673$ for $\ell=e,\mu$. If the ALPs are observed in their decay into photons, we require $L_\text{det}=1.5$\,m, such that the decay occurs before the electromagnetic calorimeter. For a given value of the Wilson coefficients $C_{Zh}$ or $C_{ah}$, we can now present the reach of high-luminosity LHC searches for $h\to Za\to\ell^+\ell^- \gamma\gamma$ and $h\to aa\to 4\gamma$ decays in the $m_a-|C_{\gamma\gamma}|$ plane. We require at least 100 signal events in a dataset of 300\,fb$^{-1}$ at $\sqrt{s}=13$\,TeV (Run-2), considering gluon-fusion induced Higgs production with cross section $\sigma(pp\to h+X)=48.52$\,pb \cite{Anastasiou:2016cez} and the effective Higgs branching ratios defined above. Figure~\ref{fig:CggexclusionAA} shows this parameter space in light green. In the upper panel we present the reach of Run-2 searches for $h\to Za\to\ell^+\ell^-\gamma\gamma$ decays assuming $|C_{Zh}|/\Lambda=0.72\,\mbox{TeV}^{-1}$ (solid contour), $0.1\,\mbox{TeV}^{-1}$ (dashed contour) and $0.015\,\mbox{TeV}^{-1}$ (dotted contour). Reaching sensitivity to smaller $h\to Za$ branching ratios obtained with $|C_{Zh}|/\Lambda<0.015\,\mbox{TeV}^{-1}$ would require larger luminosity. The lower panel shows the reach of searches for $h\to aa\to 4\gamma$ decays assuming $|C_{ah}|/\Lambda^2=1\,\mbox{TeV}^{-2}$ (solid), $0.1\,\mbox{TeV}^{-2}$ (dashed) and $0.01\,\mbox{TeV}^{-2}$ (dotted). These contours are essentially independent of the $a\to\gamma\gamma$ branching ratio unless this quantity falls below certain threshold values. For $h\to Za$, one needs $\mbox{Br}(a\to\gamma\gamma)>3\cdot 10^{-4}$ (solid), 0.011 (dashed) and 0.46 (dotted). For $h\to aa$, one needs instead $\mbox{Br}(a\to\gamma\gamma)>0.006$ (solid), 0.049 (dashed) and 0.49 (dotted). It is thus possible to probe the ALP--photon coupling even if the ALP predominantly decays into other final states. The insensitivity of the contours to $\mbox{Br}(a\to\gamma\gamma)$ can be understood by considering the behavior of the quantity $f_{\rm dec}$ in (\ref{fdecay}). The contours limiting the green regions from the left arise from the region of large ALP decay length, $L_a\gg L_{\rm det}$, in which case $f_{\rm dec}\approx (\pi/2)\,L_{\rm det}/L_a\propto\Gamma(a\to X\bar X)/\mbox{Br}(a\to X\bar X)$. In this region the effective branching ratios in (\ref{eq:effBR}) become independent of $\mbox{Br}(a\to\gamma\gamma)$ and only depend on the partial rate $\Gamma(a\to X\bar X)\propto m_a^3\,C_{\gamma\gamma}^2$. On the other hand, the number of signal events inside the probed contour regions is bounded by the yield computed with $f_{\rm dec}=1$ (prompt ALP decays), and this number becomes too small if $\mbox{Br}(a\to X\bar X)$ falls below a critical value.

The red band in the panels shows the parameter region in which the $(g-2)_\mu$ anomaly can be explained. We consider only the theoretically preferred region $|c_{\mu\mu}|\ge|C_{\gamma\gamma}|$ and impose the constraint $|c_{\mu\mu}|/\Lambda\le 10\,\mbox{TeV}^{-1}$. In principle, larger values of $|C_{\gamma\gamma}|$ can also explain the anomaly. Almost the entire parameter space where the red band is not excluded by existing experiments -- the region between 30\,MeV and 60\,GeV -- can be covered by searches for exotic Higgs decays. Even if the relevant couplings $C_{Zh}$ and $C_{ah}$ are loop suppressed, large event yields in this region can be expected in Run-2.

\begin{figure}
\begin{center}
\includegraphics[width=0.45\textwidth]{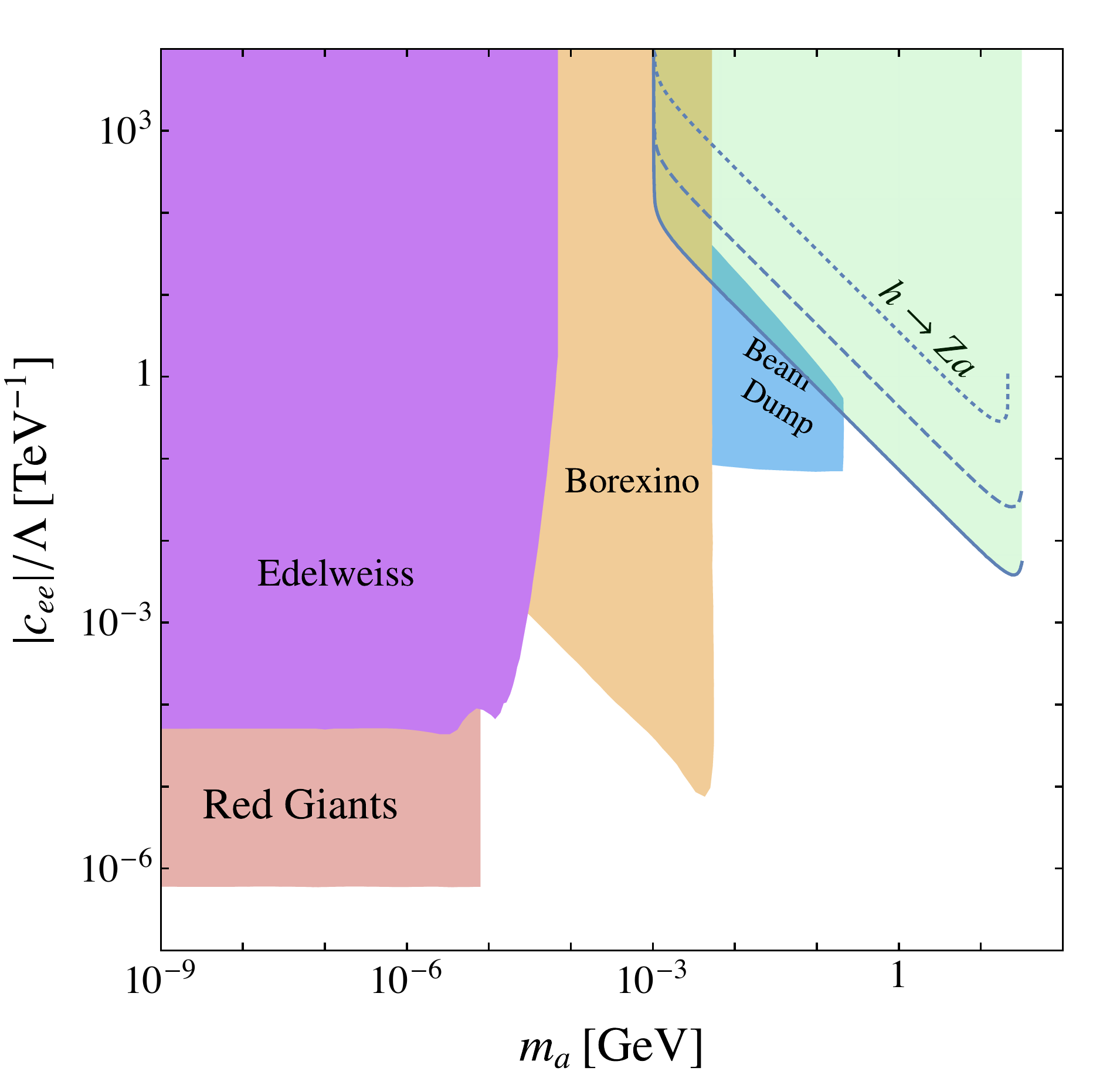}\quad\includegraphics[width=0.45\textwidth]{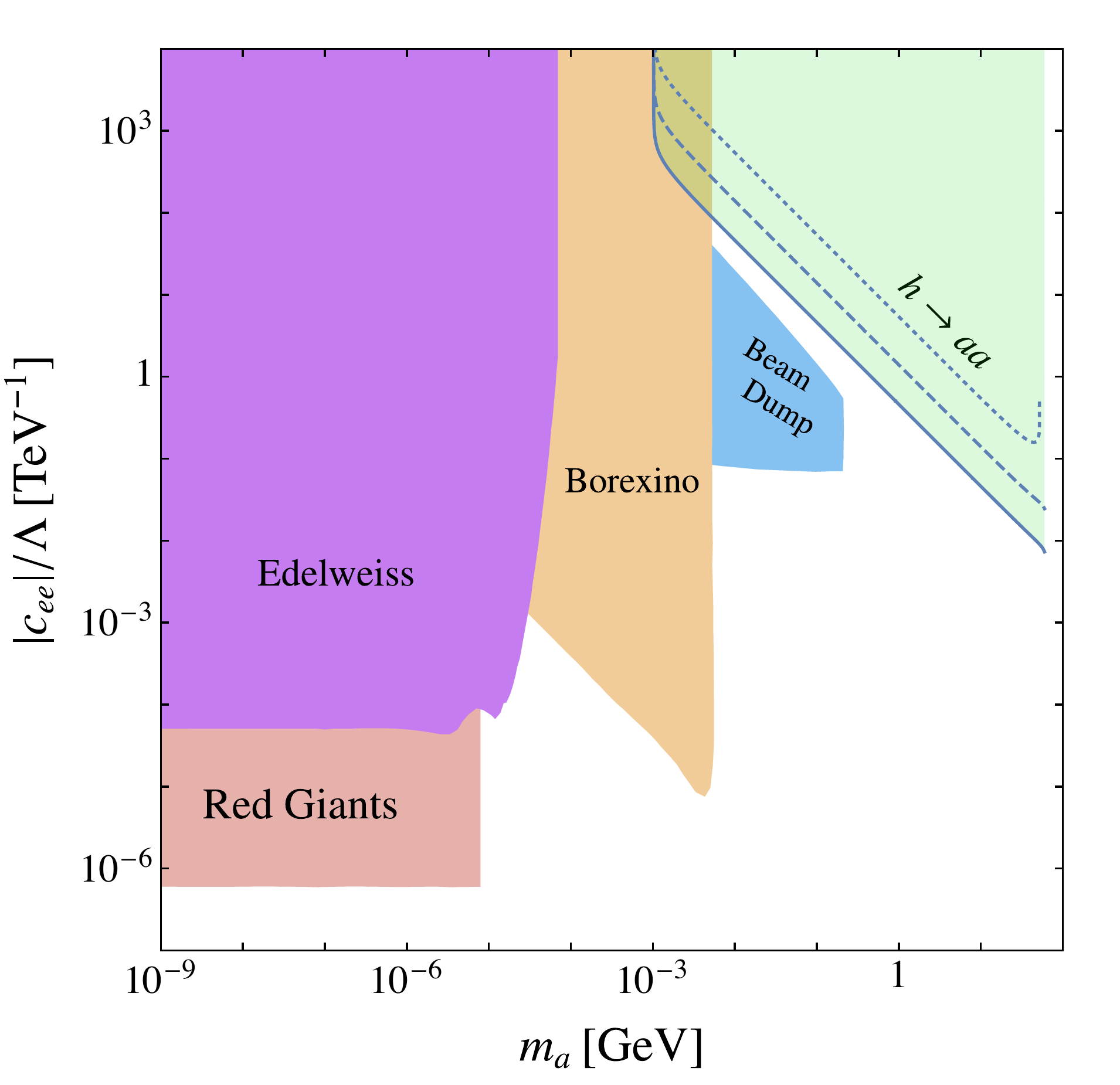}
\end{center}
\vspace{-3mm}
\caption{\label{fig:CEEexclusion} Constraints on the ALP mass and coupling to electrons derived from various experiments (colored areas without boundaries, adapted from \cite{Armengaud:2013rta,Essig:2010gu}) along with the parameter regions (shaded in light green) that can be probed in LHC Run-2 (300\,fb$^{-1}$ integrated luminosity) using the Higgs decays $h\to Za\to\ell^+\ell^-e^+e^-$ (top) and $h\to aa\to e^+e^-e^+e^-$ (bottom). We require at least 100 signal events in each channel. The contours in the upper panel correspond to $|C_{Zh}|/\Lambda=0.72\,\mbox{TeV}^{-1}$ (solid), $0.1\,\mbox{TeV}^{-1}$ (dashed) and $0.015\,\mbox{TeV}^{-1}$ (dotted). Those in the lower panel refer to $|C_{ah}|/\Lambda^2=1\,\mbox{TeV}^{-2}$ (solid), $0.1\,\mbox{TeV}^{-2}$ (dashed) and $0.01\,\mbox{TeV}^{-2}$ (dotted).}
\end{figure}

Existing searches for $h\to aa\to 4\gamma$ decay already imply interesting bounds on the ALP parameter space. ATLAS has performed dedicated searches for this signature at $m_a=100$\,MeV, 200\,MeV and 400\,MeV \cite{ATLAS:2012soa}, as well as in the high-mass region $m_a=10-62.5$\,GeV \cite{Aad:2015bua}. Lighter ALPs produced in Higgs decays would be highly boosted, and the final-state photon pairs would therefore be strongly collimated. For $m_a\lesssim 100$\,MeV these pairs cannot be resolved in the calorimeter and would be reconstructed as single photons \cite{Chang:2006bw,Dobrescu:2000jt,Draper:2012xt,Chala:2015cev}. Hence, the existing measurements of the $h\to\gamma\gamma$ rate \cite{Khachatryan:2016vau} can also be used to derive constraints on the ALP couplings. At present, non-trivial exclusion regions can be derived for values $|C_{ah}|/\Lambda^2\gtrsim 0.1\,\mbox{TeV}^{-2}$ \cite{inprep}. Currently, there exist no dedicated searches for the $h\to Za\to\ell^+\ell^-\gamma\gamma$ decay channel. However, for $m_a\lesssim 50$\,MeV the current upper bounds on the $h\to Z\gamma$ rate \cite{Chatrchyan:2013vaa,Aad:2014fia} imply a weak constraint. Since the $h\to Za$ signal does not interfere with the decay $h\to Z\gamma$, its contribution would lead to an enhancement of the $h\to Z\gamma$ rate. This would provide a very interesting signal once the decay $h\to Z\gamma$ becomes within reach of the LHC. 

The couplings of ALPs to other SM particles can be probed in an analogous way. In Figure~\ref{fig:CEEexclusion} we consider the decay $a\to e^+ e^-$. We use $L_{\rm det}=2$\,cm, such that reconstructed events correspond to decays before the inner tracker, and require 100 signal events in a dataset of 300\,fb$^{-1}$. The two panels show the reach of Run-2 searches for $h\to Za\to\ell^+\ell^-e^+e^-$ (top) and $h\to aa\to e^+e^-e^+e^-$ decays (bottom), using the same values for the Wilson coefficients $C_{Zh}$ and $C_{ah}$ as in Figure~\ref{fig:CggexclusionAA}. Once again, the contours are essentially independent of the $a\to e^+e^-$ branching ratio unless this quantity falls below certain threshold values, which are the same as before. For $h\to Za$, one needs $\mbox{Br}(a\to e^+e^-)>2\cdot 10^{-4}$ (solid), 0.011 (dashed) and 0.46 (dotted). For $h\to aa$, one needs instead $\mbox{Br}(a\to e^+e^-)>0.006$ (solid), 0.049 (dashed) and 0.49 (dotted). 

In summary, we have shown that LHC searches for the exotic Higgs decays $h\to Za$ and $h\to aa$ in Run-2 with an integrated luminosity of 300\,fb$^{-1}$ can probe the ALP couplings to photons and electrons over a large region in parameter space, which almost perfectly complements the regions covered by existing searches. Importantly, the parameter space in which an ALP can provide an explanation of the $(g-2)_\mu$ anomaly is almost completely covered by these searches. The reach can be extended with more luminosity (the event yields increase by a factor $\sim 10.7$ for 3000\,fb$^{-1}$ luminosity at $\sqrt{s}=14\,\mbox{TeV}$), and similar searches can be performed at a future lepton collider. Our yield estimates can be improved using dedicated analyses, including reconstruction efficiencies and exploiting displaced-vertex signatures. Analogous limits can also be obtained for ALP decays into pairs of muons, taus, jets, heavy quarks, as well as for invisible decays or meta-stable ALPs \cite{inprep}. 

\begin{acknowledgments}

We are grateful to Roc\'io del Rey, J\"org J\"ackel, Joachim Kopp and Pedro Schwaller for useful discussions. This work has been supported by the Cluster of Excellence {\em Precision Physics, Fundamental Interactions and Structure of Matter\/} (PRISMA -- EXC 1098) and grant 05H12UME of the German Federal Ministry for Education and Research (BMBF). 
\end{acknowledgments}


\begin{thebibliography}{99}

\bibitem{Chang:2000ii} 
  D.~Chang, W.~F.~Chang, C.~H.~Chou and W.~Y.~Keung,
  %``Large two loop contributions to g-2 from a generic pseudoscalar boson,''
  Phys.\ Rev.\ D {\bf 63}, 091301 (2001)
%  doi:10.1103/PhysRevD.63.091301
  [hep-ph/0009292].
  %%CITATION = doi:10.1103/PhysRevD.63.091301;%%
    
\bibitem{Marciano:2016yhf} 
  W.~J.~Marciano, A.~Masiero, P.~Paradisi and M.~Passera,
  %``Contributions of axionlike particles to lepton dipole moments,''
  Phys.\ Rev.\ D {\bf 94}, no. 11, 115033 (2016)
%  doi:10.1103/PhysRevD.94.115033
  [arXiv:1607.01022 [hep-ph]].
  %%CITATION = doi:10.1103/PhysRevD.94.115033;%%

\bibitem{Kleban:2005rj} 
  M.~Kleban and R.~Rabadan,
  %``Collider bounds on pseudoscalars coupling to gauge bosons,''
  hep-ph/0510183.
  %%CITATION = HEP-PH/0510183;%%

\bibitem{Mimasu:2014nea} 
  K.~Mimasu and V.~Sanz,
  %``ALPs at Colliders,''
  JHEP {\bf 1506}, 173 (2015)
%  doi:10.1007/JHEP06(2015)173
  [arXiv:1409.4792 [hep-ph]].
  %%CITATION = doi:10.1007/JHEP06(2015)173;%%

\bibitem{Jaeckel:2015jla} 
  J.~Jaeckel and M.~Spannowsky,
  %``Probing MeV to 90 GeV axion-like particles with LEP and LHC,''
  Phys.\ Lett.\ B {\bf 753}, 482 (2016)
%  doi:10.1016/j.physletb.2015.12.037
  [arXiv:1509.00476 [hep-ph]].
  %%CITATION = doi:10.1016/j.physletb.2015.12.037;%%

\bibitem{Brivio:2017ije} 
%  I.~Brivio, M.~B.~Gavela, L.~Merlo, K.~Mimasu, J.~M.~No, R.~del Rey and V.~Sanz,
  I.~Brivio {\it et al.},
  %``ALPs Effective Field Theory and Collider Signatures,''
  arXiv:1701.05379 [hep-ph].
  %%CITATION = ARXIV:1701.05379;%%

\bibitem{Kim:1989xj} 
  J.~E.~Kim and U.~W.~Lee,
  %``$\Z^0$ Decay to Photon and Variant Axion,''
  Phys.\ Lett.\ B {\bf 233}, 496 (1989).
%  doi:10.1016/0370-2693(89)91347-6
  %%CITATION = doi:10.1016/0370-2693(89)91347-6;%%
  
\bibitem{Djouadi:1990ms} 
  A.~Djouadi, P.~M.~Zerwas and J.~Zunft,
  %``Search for light pseudoscalar Higgs bosons in Z decays,''
  Phys.\ Lett.\ B {\bf 259}, 175 (1991).
%  doi:10.1016/0370-2693(91)90155-J
  %%CITATION = doi:10.1016/0370-2693(91)90155-J;%%

\bibitem{Rupak:1995kg} 
  G.~Rupak and E.~H.~Simmons,
  %``Limits on pseudoscalar bosons from rare Z decays at LEP,''
  Phys.\ Lett.\ B {\bf 362}, 155 (1995)
%  doi:10.1016/0370-2693(95)01152-G
  [hep-ph/9507438].
  %%CITATION = doi:10.1016/0370-2693(95)01152-G;%%

\bibitem{Dobrescu:2000jt} 
  B.~A.~Dobrescu, G.~L.~Landsberg and K.~T.~Matchev,
  %``Higgs boson decays to CP odd scalars at the Tevatron and beyond,''
  Phys.\ Rev.\ D {\bf 63}, 075003 (2001)
%  doi:10.1103/PhysRevD.63.075003
  [hep-ph/0005308].
  %%CITATION = doi:10.1103/PhysRevD.63.075003;%%

\bibitem{Dobrescu:2000yn} 
  B.~A.~Dobrescu and K.~T.~Matchev,
  %``Light axion within the next-to-minimal supersymmetric standard model,''
  JHEP {\bf 0009}, 031 (2000)
%  doi:10.1088/1126-6708/2000/09/031
  [hep-ph/0008192].
  %%CITATION = doi:10.1088/1126-6708/2000/09/031;%%

\bibitem{Chang:2006bw} 
  S.~Chang, P.~J.~Fox and N.~Weiner,
  %``Visible Cascade Higgs Decays to Four Photons at Hadron Colliders,''
  Phys.\ Rev.\ Lett.\  {\bf 98}, 111802 (2007)
%  doi:10.1103/PhysRevLett.98.111802
  [hep-ph/0608310].
  %%CITATION = doi:10.1103/PhysRevLett.98.111802;%%
 
\bibitem{Chatrchyan:2012cg} 
  S.~Chatrchyan {\it et al.} [CMS Collaboration],
  %``Search for a non-standard-model Higgs boson decaying to a pair of new light bosons in four-muon final states,''
  Phys.\ Lett.\ B {\bf 726}, 564 (2013)
%  doi:10.1016/j.physletb.2013.09.009
  [arXiv:1210.7619 [hep-ex]].
  %%CITATION = doi:10.1016/j.physletb.2013.09.009;%%

\bibitem{CMS:2015iga} 
  CMS Collaboration,
  %``Search for Higgs Decays to New Light Bosons in Boosted Tau Final States,''
  CMS-PAS-HIG-14-022.
  %%CITATION = CMS-PAS-HIG-14-022;%%

\bibitem{CMS:2016cel} 
  CMS Collaboration,
  %``Search for exotic decays of the Higgs boson to a pair of new light bosons with two muon and two b jets in final states,''
  CMS-PAS-HIG-14-041.
  %%CITATION = CMS-PAS-HIG-14-041;%%

\bibitem{Aad:2015bua} 
  G.~Aad {\it et al.} [ATLAS Collaboration],
  %``Search for new phenomena in events with at least three photons collected in $pp$ collisions at $\sqrt{s}$ = 8 TeV with the ATLAS detector,''
  Eur.\ Phys.\ J.\ C {\bf 76}, no. 4, 210 (2016)
%  doi:10.1140/epjc/s10052-016-4034-8
  [arXiv:1509.05051 [hep-ex]].
  %%CITATION = doi:10.1140/epjc/s10052-016-4034-8;%%

\bibitem{Khachatryan:2015nba} 
  V.~Khachatryan {\it et al.} [CMS Collaboration],
  %``Search for a very light NMSSM Higgs boson produced in decays of the 125 GeV scalar boson and decaying into $\tau$ leptons in pp collisions at $\sqrt{s}=8$ TeV,''
  JHEP {\bf 1601}, 079 (2016)
%  doi:10.1007/JHEP01(2016)079
  [arXiv:1510.06534 [hep-ex]].
  %%CITATION = doi:10.1007/JHEP01(2016)079;%%

\bibitem{CMS:2016tgd} 
  CMS Collaboration,
  %``A Search for Beyond Standard Model Light Bosons Decaying into Muon Pairs,''
  CMS-PAS-HIG-16-035.
  %%CITATION = CMS-PAS-HIG-16-035;%%
  
\bibitem{Khachatryan:2017mnf} 
  V.~Khachatryan {\it et al.} [CMS Collaboration],
  %``Search for light bosons in decays of the 125 GeV Higgs boson in proton-proton collisions at $\sqrt{s}$ = 8 TeV,''
  arXiv:1701.02032 [hep-ex].
  %%CITATION = ARXIV:1701.02032;%%

\bibitem{Khachatryan:2016are} 
  V.~Khachatryan {\it et al.} [CMS Collaboration],
  %``Search for neutral resonances decaying into a Z boson and a pair of b jets or tau leptons,''
  Phys.\ Lett.\ B {\bf 759}, 369 (2016)
%  doi:10.1016/j.physletb.2016.05.087
  [arXiv:1603.02991 [hep-ex]].
  %%CITATION = doi:10.1016/j.physletb.2016.05.087;%%

\bibitem{Branco:2011iw} 
%  G.~C.~Branco, P.~M.~Ferreira, L.~Lavoura, M.~N.~Rebelo, M.~Sher and J.~P.~Silva,
  G.~C.~Branco {\it et al.},
  %``Theory and phenomenology of two-Higgs-doublet models,''
  Phys.\ Rept.\  {\bf 516}, 1 (2012)
%  doi:10.1016/j.physrep.2012.02.002
  [arXiv:1106.0034 [hep-ph]].
  %%CITATION = doi:10.1016/j.physrep.2012.02.002;%%

\bibitem{Bauer:2016ydr} 
  M.~Bauer, M.~Neubert and A.~Thamm,
  %``The "forgotten" decay S -> Z+h as a CP analyzer,''
  arXiv:1607.01016 [hep-ph];
  %%CITATION = ARXIV:1607.01016;%%
%
%\bibitem{Bauer:2016zfj} 
%  M.~Bauer, M.~Neubert and A.~Thamm,
  %``Analyzing the CP Nature of a New Scalar Particle via S->Zh Decay,''
  Phys.\ Rev.\ Lett.\  {\bf 117}, 181801 (2016)
%  doi:10.1103/PhysRevLett.117.181801
  [arXiv:1610.00009 [hep-ph]].
  %%CITATION = doi:10.1103/PhysRevLett.117.181801;%%
  
\bibitem{inprep} 
  M.~Bauer, M.~Neubert and A.~Thamm,
  %``Analyzing the CP Nature of a New Scalar Particle via S->Zh Decay,''
  in preparation.

\bibitem{Georgi:1986df} 
  H.~Georgi, D.~B.~Kaplan and L.~Randall,
  %``Manifesting the Invisible Axion at Low-energies,''
  Phys.\ Lett.\  {\bf 169B}, 73 (1986).
%  doi:10.1016/0370-2693(86)90688-X
  %%CITATION = doi:10.1016/0370-2693(86)90688-X;%%

\bibitem{Olive:2016xmw} 
  C.~Patrignani {\it et al.} [Particle Data Group],
  %``Review of Particle Physics,''
  Chin.\ Phys.\ C {\bf 40}, no. 10, 100001 (2016).
%  doi:10.1088/1674-1137/40/10/100001
  %%CITATION = doi:10.1088/1674-1137/40/10/100001;%%
  
\bibitem{Leveille:1977rc} 
  J.~P.~Leveille,
  %``The Second Order Weak Correction to (G-2) of the Muon in Arbitrary Gauge Models,''
  Nucl.\ Phys.\ B {\bf 137}, 63 (1978).
%  doi:10.1016/0550-3213(78)90051-2
  %%CITATION = doi:10.1016/0550-3213(78)90051-2;%%
  
\bibitem{Haber:1978jt} 
  H.~E.~Haber, G.~L.~Kane and T.~Sterling,
  %``The Fermion Mass Scale and Possible Effects of Higgs Bosons on Experimental Observables,''
  Nucl.\ Phys.\ B {\bf 161}, 493 (1979).
%  doi:10.1016/0550-3213(79)90225-6
  %%CITATION = doi:10.1016/0550-3213(79)90225-6;%%

\bibitem{Feruglio:1992wf} 
  F.~Feruglio,
  %``The Chiral approach to the electroweak interactions,''
  Int.\ J.\ Mod.\ Phys.\ A {\bf 8}, 4937 (1993)
%  doi:10.1142/S0217751X93001946
  [hep-ph/9301281].
  %%CITATION = doi:10.1142/S0217751X93001946;%%

\bibitem{Khachatryan:2016vau} 
  G.~Aad {\it et al.} [ATLAS and CMS Collaborations],
  %``Measurements of the Higgs boson production and decay rates and constraints on its couplings from a combined ATLAS and CMS analysis of the LHC pp collision data at $ \sqrt{s}=7 $ and 8 TeV,''
  JHEP {\bf 1608}, 045 (2016)
%  doi:10.1007/JHEP08(2016)045
  [arXiv:1606.02266 [hep-ex]].
  %%CITATION = doi:10.1007/JHEP08(2016)045;%%

\bibitem{Anastasiou:2016cez} 
%  C.~Anastasiou, C.~Duhr, F.~Dulat, E.~Furlan, T.~Gehrmann, F.~Herzog, A.~Lazopoulos and B.~Mistlberger,
  C.~Anastasiou {\it et al.},
  %``High precision determination of the gluon fusion Higgs boson cross-section at the LHC,''
  JHEP {\bf 1605}, 058 (2016)
%  doi:10.1007/JHEP05(2016)058
  [arXiv:1602.00695 [hep-ph]].
  %%CITATION = doi:10.1007/JHEP05(2016)058;%%
  
\bibitem{ATLAS:2012soa} 
  ATLAS Collaboration,
  %``Search for a Higgs boson decaying to four photons through light CP-odd scalar coupling using \lumifull of $7~\mathrm{TeV}$ $pp$ collision data taken with ATLAS detector at the LHC,''
  ATLAS-CONF-2012-079.
  %%CITATION = ATLAS-CONF-2012-079;%%

\bibitem{Draper:2012xt} 
  P.~Draper and D.~McKeen,
  %``Diphotons from Tetraphotons in the Decay of a 125 GeV Higgs at the LHC,''
  Phys.\ Rev.\ D {\bf 85}, 115023 (2012)
%  doi:10.1103/PhysRevD.85.115023
  [arXiv:1204.1061 [hep-ph]].
  %%CITATION = doi:10.1103/PhysRevD.85.115023;%%

\bibitem{Chala:2015cev} 
  M.~Chala, M.~Duerr, F.~Kahlhoefer and K.~Schmidt-Hoberg,
  %``Tricking Landau?Yang: How to obtain the diphoton excess from a vector resonance,''
  Phys.\ Lett.\ B {\bf 755}, 145 (2016)
%  doi:10.1016/j.physletb.2016.02.006
  [arXiv:1512.06833 [hep-ph]].
  %%CITATION = doi:10.1016/j.physletb.2016.02.006;%%

\bibitem{Chatrchyan:2013vaa} 
  S.~Chatrchyan {\it et al.} [CMS Collaboration],
  %``Search for a Higgs boson decaying into a Z and a photon in pp collisions at sqrt(s) = 7 and 8 TeV,''
  Phys.\ Lett.\ B {\bf 726}, 587 (2013)
%  doi:10.1016/j.physletb.2013.09.057
  [arXiv:1307.5515 [hep-ex]].
  %%CITATION = doi:10.1016/j.physletb.2013.09.057;%%
  
\bibitem{Aad:2014fia} 
  G.~Aad {\it et al.} [ATLAS Collaboration],
  %``Search for Higgs boson decays to a photon and a Z boson in pp collisions at $\sqrt{s}$=7 and 8 TeV with the ATLAS detector,''
  Phys.\ Lett.\ B {\bf 732}, 8 (2014)
%  doi:10.1016/j.physletb.2014.03.015
  [arXiv:1402.3051 [hep-ex]].
  %%CITATION = doi:10.1016/j.physletb.2014.03.015;%%

\bibitem{Armengaud:2013rta}
 E.~Armengaud {\it et al.},
 %``Axion searches with the EDELWEISS-II experiment,''
 JCAP {\bf 1311}, 067 (2013)
% doi:10.1088/1475-7516/2013/11/067
% [arXiv:1307.1488 [astro-ph.CO]].
 [arXiv:1307.1488 [astro-ph]].

\bibitem{Essig:2010gu} 
  R.~Essig, R.~Harnik, J.~Kaplan and N.~Toro,
  %``Discovering New Light States at Neutrino Experiments,''
  Phys.\ Rev.\ D {\bf 82}, 113008 (2010)
%  doi:10.1103/PhysRevD.82.113008
  [arXiv:1008.0636 [hep-ph]].
  %%CITATION = doi:10.1103/PhysRevD.82.113008;%%

\end{thebibliography}
\end{document}